\documentclass[dvips]{article}
\usepackage{supertabular,lscape,epsfig}
\usepackage{amssymb}
\usepackage{amsmath}
\DeclareSymbolFont{ppa}{OT1}{ppl}{m}{it}
\DeclareMathSymbol{\vv}{\mathalpha}{ppa}{'166}

\begin{document}

\newcommand{\dd}{\,{\rm d}}
\newcommand{\ie}{{\it i.e.},\,}
\newcommand{\etal}{{\it et al.\ }}
\newcommand{\eg}{{\it e.g.},\,}
\newcommand{\cf}{{\it cf.\ }}
\newcommand{\vs}{{\it vs.\ }}
\newcommand{\zdot}{\makebox[0pt][l]{.}}
\newcommand{\up}[1]{\ifmmode^{\rm #1}\else$^{\rm #1}$\fi}
\newcommand{\dn}[1]{\ifmmode_{\rm #1}\else$_{\rm #1}$\fi}
\newcommand{\upd}{\up{d}}
\newcommand{\uph}{\up{h}}
\newcommand{\upm}{\up{m}}  
\newcommand{\ups}{\up{s}}
\newcommand{\arcd}{\ifmmode^{\circ}\else$^{\circ}$\fi}
\newcommand{\arcm}{\ifmmode{'}\else$'$\fi}
\newcommand{\arcs}{\ifmmode{''}\else$''$\fi}
\newcommand{\MS}{{\rm M}\ifmmode_{\odot}\else$_{\odot}$\fi}
\newcommand{\RS}{{\rm R}\ifmmode_{\odot}\else$_{\odot}$\fi}
\newcommand{\LS}{{\rm L}\ifmmode_{\odot}\else$_{\odot}$\fi}

\newcommand{\Abstract}[2]{{\footnotesize\begin{center}ABSTRACT\end{center}
\vspace{1mm}\par#1\par   
\noindent
{~}{\it #2}}}

\newcommand{\TabCap}[2]{\begin{center}\parbox[t]{#1}{\begin{center}
  \small {\spaceskip 2pt plus 1pt minus 1pt T a b l e}
  \refstepcounter{table}\thetable \\[2mm]
  \footnotesize #2 \end{center}}\end{center}}

\newcommand{\TableSep}[2]{\begin{table}[p]\vspace{#1}
\TabCap{#2}\end{table}}

\newcommand{\FigCap}[1]{\footnotesize\par\noindent Fig.\  %
  \refstepcounter{figure}\thefigure. #1\par}

\newcommand{\TableFont}{\footnotesize}
\newcommand{\TableFontIt}{\ttit}
\newcommand{\SetTableFont}[1]{\renewcommand{\TableFont}{#1}}

\newcommand{\MakeTable}[4]{\begin{table}[htb]\TabCap{#2}{#3}
  \begin{center} \TableFont \begin{tabular}{#1} #4
  \end{tabular}\end{center}\end{table}}

\newcommand{\MakeTableSep}[4]{\begin{table}[p]\TabCap{#2}{#3}
  \begin{center} \TableFont \begin{tabular}{#1} #4
  \end{tabular}\end{center}\end{table}}

\newenvironment{references}%
{
\footnotesize \frenchspacing
\renewcommand{\thesection}{}
\renewcommand{\in}{{\rm in }}
\renewcommand{\AA}{Astron.\ Astrophys.}
\newcommand{\AAS}{Astron.~Astrophys.~Suppl.~Ser.}
\newcommand{\ApJ}{Astrophys.\ J.}
\newcommand{\ApJS}{Astrophys.\ J.~Suppl.~Ser.}
\newcommand{\ApJL}{Astrophys.\ J.~Letters}
\newcommand{\AJ}{Astron.\ J.}
\newcommand{\IBVS}{IBVS}
\newcommand{\PASP}{P.A.S.P.}
\newcommand{\Acta}{Acta Astron.}
\newcommand{\MNRAS}{MNRAS}
\renewcommand{\and}{{\rm and }}
\section{{\rm REFERENCES}}
\sloppy \hyphenpenalty10000
\begin{list}{}{\leftmargin1cm\listparindent-1cm
\itemindent\listparindent\parsep0pt\itemsep0pt}}%
{\end{list}\vspace{2mm}}
 
\def\TYLDA{~}
\newlength{\DW}
\settowidth{\DW}{0}
\newcommand{\dw}{\hspace{\DW}}

\newcommand{\refitem}[5]{\item[]{#1} #2%
\def\REFARG{#3}\ifx\REFARG\TYLDA\else, {\it#3}\fi
\def\REFARG{#4}\ifx\REFARG\TYLDA\else, {\bf#4}\fi
\def\REFARG{#5}\ifx\REFARG\TYLDA\else, {#5}\fi.}

\newcommand{\Section}[1]{\section{#1}}
\newcommand{\Subsection}[1]{\subsection{#1}}
\newcommand{\Acknow}[1]{\par\vspace{5mm}{\bf Acknowledgements.} #1}
\pagestyle{myheadings}

\newfont{\bb}{ptmbi8t at 12pt}
\newcommand{\xrule}{\rule{0pt}{2.5ex}}  
\newcommand{\xxrule}{\rule[-1.8ex]{0pt}{4.5ex}}  
\def\thefootnote{\fnsymbol{footnote}}
\begin{center}

{\Large\bf
Cluster AgeS Experiment (CASE): Dwarf Novae and a Probable
Microlensing Event in the Globular Cluster M22}
\vskip1cm
{\bf
P.~~P~i~e~t~r~u~k~o~w~i~c~z$^1$,~~ J.~~~K~a~l~u~z~n~y$^1$,
~~I.~B.~~T~h~o~m~p~s~o~n$^2$,\\
~~M.~~J~a~r~o~s~z~y~\'n~s~k~i$^3$,
~~A.~~S~c~h~w~a~r~z~e~n~b~e~r~g-C~z~e~r~n~y$^{1,4}$,\\
~~W.~~K~r~z~e~m~i~n~s~k~i$^5$ and~~ W.~~P~y~c~h$^1$\\}
\vskip3mm
{
  $^1$Nicolaus Copernicus Astronomical Center,
     ul. Bartycka 18, 00-716 Warsaw, Poland\\
     e-mail: (pietruk,jka,alex,pych)\\
  $^2$Carnegie Institution of Washington,
     813 Santa Barbara Street, Pasadena, CA 91101, USA\\
     e-mail: ian@ociw.edu\\
  $^3$Warsaw University Observatory,
     Al. Ujazdowskie 4, 00-478 Warsaw, Poland\\
     e-mail: mj@astrouw.edu.pl\\
  $^4$Astronomical Observatory, Adam Mickiewicz University,
     ul. S{\l}oneczna 36, 61-286 Pozna\'n, Poland\\
  $^5$Las Campanas Observatory, Casilla 601, La Serena, Chile,\\
     e-mail: wojtek@lco.cl}
\end{center}

\Abstract{

We report the identification of a new cataclysmic variable (denoted as
CV2) and a probable microlensing event in the field of the globular
cluster M22. Two outbursts were observed for CV2. During one of them
superhumps with $P_{sh}=0.08875$~d were present in the light curve.
CV2 has an X-ray counterpart detected by XMM-Newton. A very likely
microlensing event at a radius of $2\zdot\arcm3$ from the cluster center
was detected. It had an amplitude of $\Delta V=0.75$~mag and a
characteristic time of 15.9~days. Based on model considerations we
show that the most likely configuration has the source in the Galactic
bulge with the lens in the cluster. Two outbursts were observed for the
already known dwarf nova CV1.
}
{Stars: dwarf novae -- novae, cataclysmic variables -- globular
clusters: individual: (M22, NGC~6656) -- Gravitational lensing}

\Section{Introduction}

The dense central regions of globular clusters (GCs) offer a unique
opportunity for studies of dynamical processes in stellar systems.
It is well established that even a small 
population of close binaries can drive the evolution of the entire 
cluster (Hut \etal 1992, 2003). It is expected on theoretical 
grounds that large numbers of cataclysmic variables (CVs) should be 
present in the central parts of GCs (Fabian \etal 1975; 
Hut and Paczy\'nski 1984). 
Several tens of candidate cluster CVs have been reported over the last    
few years based on observations collected with the Chandra
(\eg Hannikainen \etal 2005, Heinke \etal 2005) and XMM-Newton
(\eg Gendre \etal 2003; Webb \etal 2004) telescopes. However, 
only a few CVs in GCs have been confirmed with spectroscopic observations.
Similarly, there are only a few objects for which dwarf nova (DN) 
type outbursts have been observed (see Kaluzny \etal 2005 
for more extended summary of the subject).

This contribution continues a series of papers devoted to the search
for CVs in GCs based on the rich photometric data base collected by the 
CASE collaboration (Kaluzny \etal 2005). We report here the results for the 
cluster M22 (NGC~6656) which is located about
one-third of the way between the Sun and the Galactic bulge.
Two DN type outbursts were detected for the previously known variable CV1 and
for the newly identified variable CV2. We have also observed
a very likely microlensing event caused by a cluster star. 

\Section{Observations and Search for Erupting Objects}

The CASE project is conducted at Las Campanas Observatory. For the
survey we used the 1.0-m Swope telescope equipped with a
$2048\times 3150$ pixel SITE3 CCD camera. With a scale of 0.435 
arcsec/pixel the field of view was $14.8\times 23$~arcmin$^{2}$.
A fraction of the images of M22 was taken 
in a subrastered mode with a field of $14.8\times 15.6$~arcmin$^{2}$.
The analysis presented in this paper is based on a uniform set of 
trimmed images with a field of view $14.8\times 11.6$~arcmin$^{2}$
(longer axis along E-W direction). The cluster core was located 
approximately at the center of the images.

The cluster was monitored during the 2000 and 2001 seasons.
A total of 2006 images were taken through the $V$ filter
with exposure times 90-300s. In addition,
we have obtained 390 images in the $B$ filter with exposure times
140-300s. The number of $V$-band frames taken per night ranged from
a few to 62. The median seeing was $1\zdot\arcs41$ and $1\zdot\arcs43$
for the $V$ and $B$ bands, respectively.

The photometry was extracted with the help of the {\it Difference Image
Analysis Package} (DIAPL) written by Wo\'zniak (2000) and recently modified
by W. Pych. The package is an implementation of the 
method developed by Alard and Lupton (1998).
The procedure used to search for possible "outbursting" objects 
is described in detail in Kaluzny \etal (2005). The search was based on 
the $V$ band data collected on 71 nights for which at least one
image with seeing better than $1\zdot\arcs6$ was available.
A reference frame for the $V$ filter was constructed by combining 19 individual
frames taken on the night of 2000 Sep 2/3. The seeing for that frame
was $FWHM=1\zdot\arcs03$. Profile photometry, 
as well as aperture corrections for the reference images, was
extracted with DAOPHOT/ALLSTAR (Stetson 1987). These
measurements were subsequently 
used to transform the light curves from differential flux units 
into instrumental magnitudes. The final transformations onto the standard
$BV$ system were based on observations of several Landolt (1992)
fields which were collected on one of the nights 
during which M22 was observed.    

Searches for erupting objects in M22 led to the detection of 
three objects. In addition to the previously known DN CV1
(Anderson \etal 2003; Bond \etal 2005), we identified another DN 
(which we provisionally name CV2) and a probable microlensing event. 
The equatorial coordinates of these three objects, as well as their
angular distances from the cluster center
($\alpha_{2000}=18^h36^m24\zdot\ups2$,
$\delta_{2000}=-23^{\arcd}54^{\arcm}12^{\arcs}$, Harris 1996),
are listed in Table 1.

\begin{table}
\centering
\caption{\small Positions of eruptive objects observed in M22}
{\small
\begin{tabular}{lccc}
\hline
Name & RA(2000.0) & Dec(2000.0) & Distance \\
 & & & from center \\
\hline
M22~CV1       & 18\uph36\upm24\zdot\ups66 & -23\arcd54\arcm35\zdot\arcs5 & 0\zdot\arcm40 \\
M22~CV2       & 18\uph36\upm02\zdot\ups72 & -23\arcd55\arcm24\zdot\arcs6 & 5\zdot\arcm51 \\
M22~microlens & 18\uph36\upm22\zdot\ups40 & -23\arcd56\arcm29\zdot\arcs4 & 2\zdot\arcm33 \\
\hline
\end{tabular}}
\end{table}

\Section{Outbursts of CV1}

The variability of CV1 was first noted by Sahu \etal (2001). They 
interpreted the May 1999 brightening episode detected 
on HST/WFPC2 images as a microlensing event. Subsequently
Anderson \etal (2003)
argued that in fact the variable is a CV and that the 
observed event was due to a DN type outburst.
The outburst lasted about 25 days with an approximate amplitude 
of 3 mag, peaking at $I\approx15$ mag. Two other outburst episodes, 
one in 2002 and one in 2003, were reported by Bond \etal (2005). For the event
in 2002 only one exposure was obtained showing the variable at $I=15.1$.
During the 2003 episode the variable was observed at 
$I_{max}\approx15.7$ mag, but the actual peak 
of the outburst was very likely not observed.

\begin{figure}[htb]
\centerline{\includegraphics[height=120mm,width=120mm]{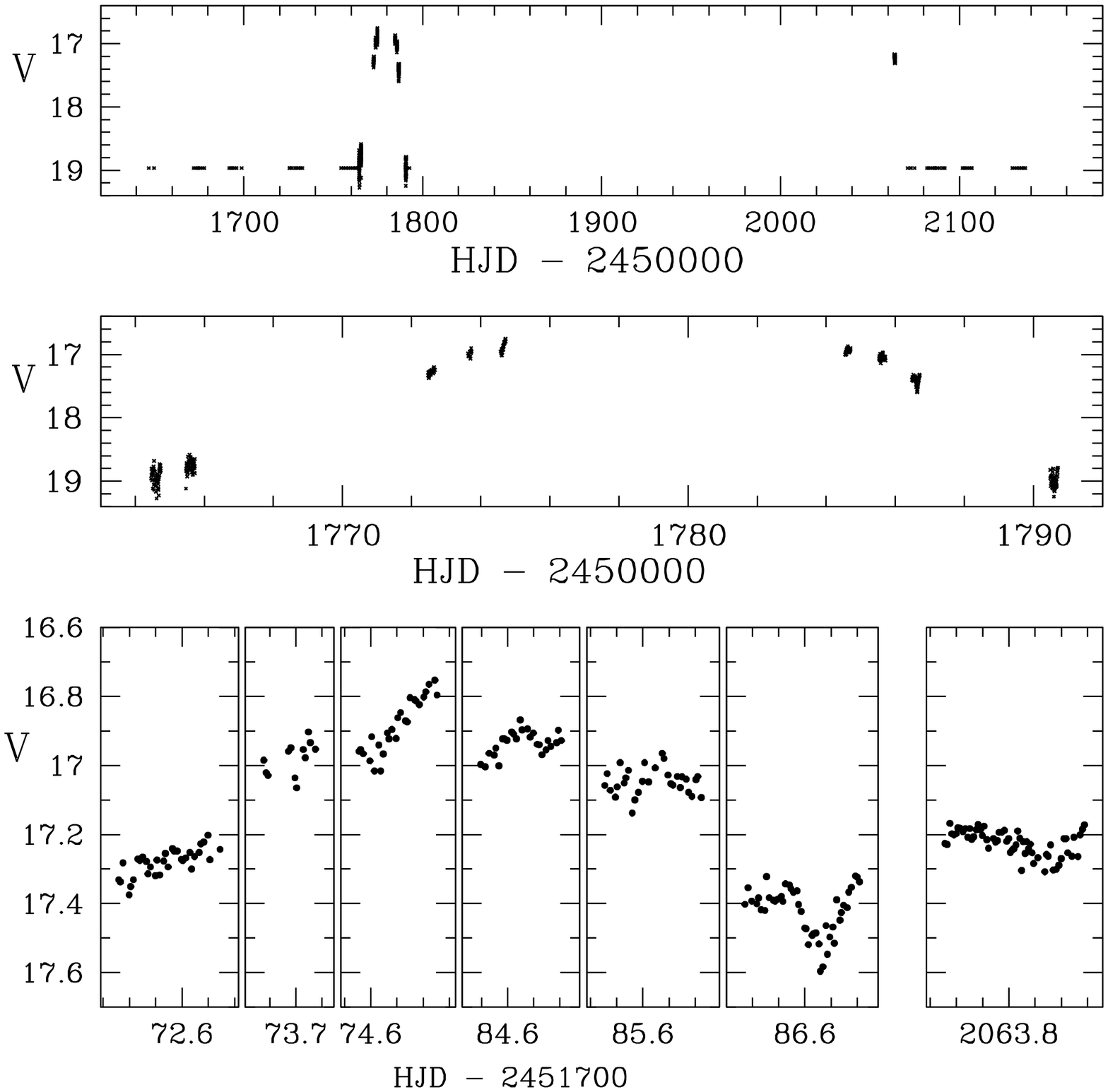}}
\FigCap{
Light curves of CV1: full span in the 2000-2001 seasons
(upper panel), during the 2000 Aug/Sep outburst (middle panel)
and nightly segmented during the outbursts (lower panels).
In the segments small ticks are every 0.05 days.
}
\end{figure}

As can be seen in Fig. 1 another two outbursts of CV1 were 
detected in our data. The first one occurred on  
August/September 2000 and lasted approximately 20 days. 
The observed maximum reached $V=16.75$ mag, but we failed
to cover the actual peak of the outburst. There is no
evidence for the presence of superhumps in the light curve. 
The second outburst episode of CV1 in our data set occurred in 2001.
On June 3, 2001 the variable had a mean magnitude of $V=17.22$.
Unfortunately, we observed the cluster on only one night
during this event, and observations collected 
seven days later showed that CV1 had already faded to a low state. 

Our data taken together with the data
of Bond \etal (2005) cover the observing seasons 2000-2004.
A total of 4 outbursts were observed during that period suggesting 
that the average recurrence time is rather long, 
probably exceeding 150 days. The duration of the three outbursts 
with reasonably complete time coverage ranges from 
about 20 to about 25 days (Sahu \etal 2001; Bond 2005; this paper).

In Fig. 2 the location of CV1 during the 2000 eruption is marked on a
color-magnitude diagram of M22. The cluster main sequence shows a
noticeable spread due to the presence of significant differential
reddening ($0.34<E(B-V)<0.42$ (Richter \etal 1999)). 
In the following analysis we use $E(B-V)=0.38\pm 0.04$ for the 
cluster reddening.
The variable had a mean color of $<B-V>=0.46 \pm 0.09$ near maximum light
which implies an unreddened color of $(B-V)_0=0.08 \pm 0.13$. This is
within the range of colors exhibited by DNe during outbursts (Warner
1995). We were unable to obtain reliable photometry of the variable
on many images collected in the low state due to crowding effects. From
archival HST/WFPC2 images (GO 8174 program) observed on 2000 June 28 we
obtained $V_{min}=18.96$. The photometry was extracted with the help of
HSTphot package (Dolphin 2000a,b). In Fig. 1 we adopt this value
of $V_{min}$ to mark all out-of-outburst observations of CV1.

\begin{figure}[htb]
\centerline{\includegraphics[height=120mm,width=120mm]{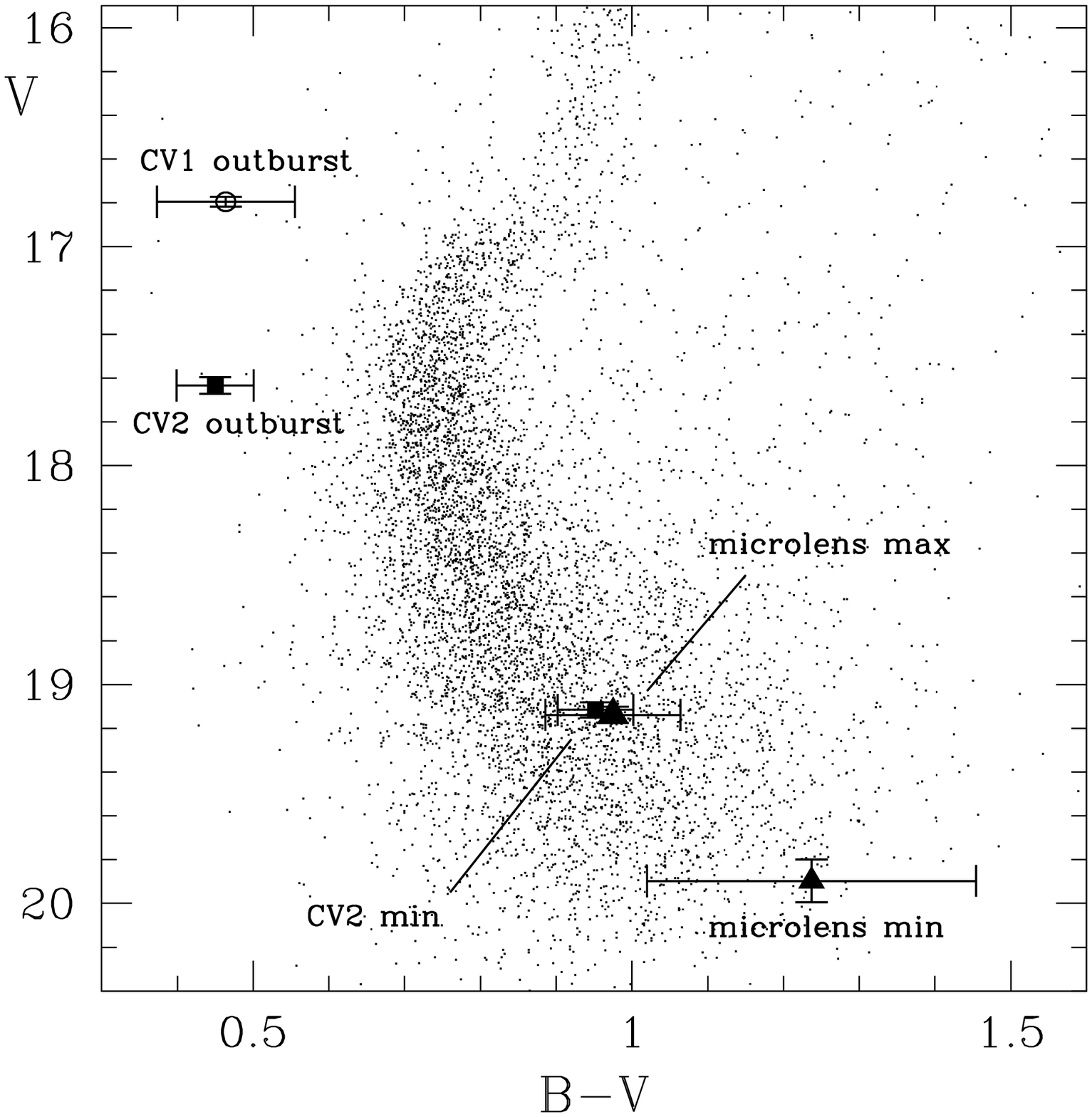}}
\FigCap{
Locations of observed cataclysmic variables
and the microlensing event on the color-magnitude diagram of M22
}
\end{figure}

\Section{Variable CV2}

\begin{figure}[!b]
\centerline{\includegraphics[height=120mm,width=120mm]{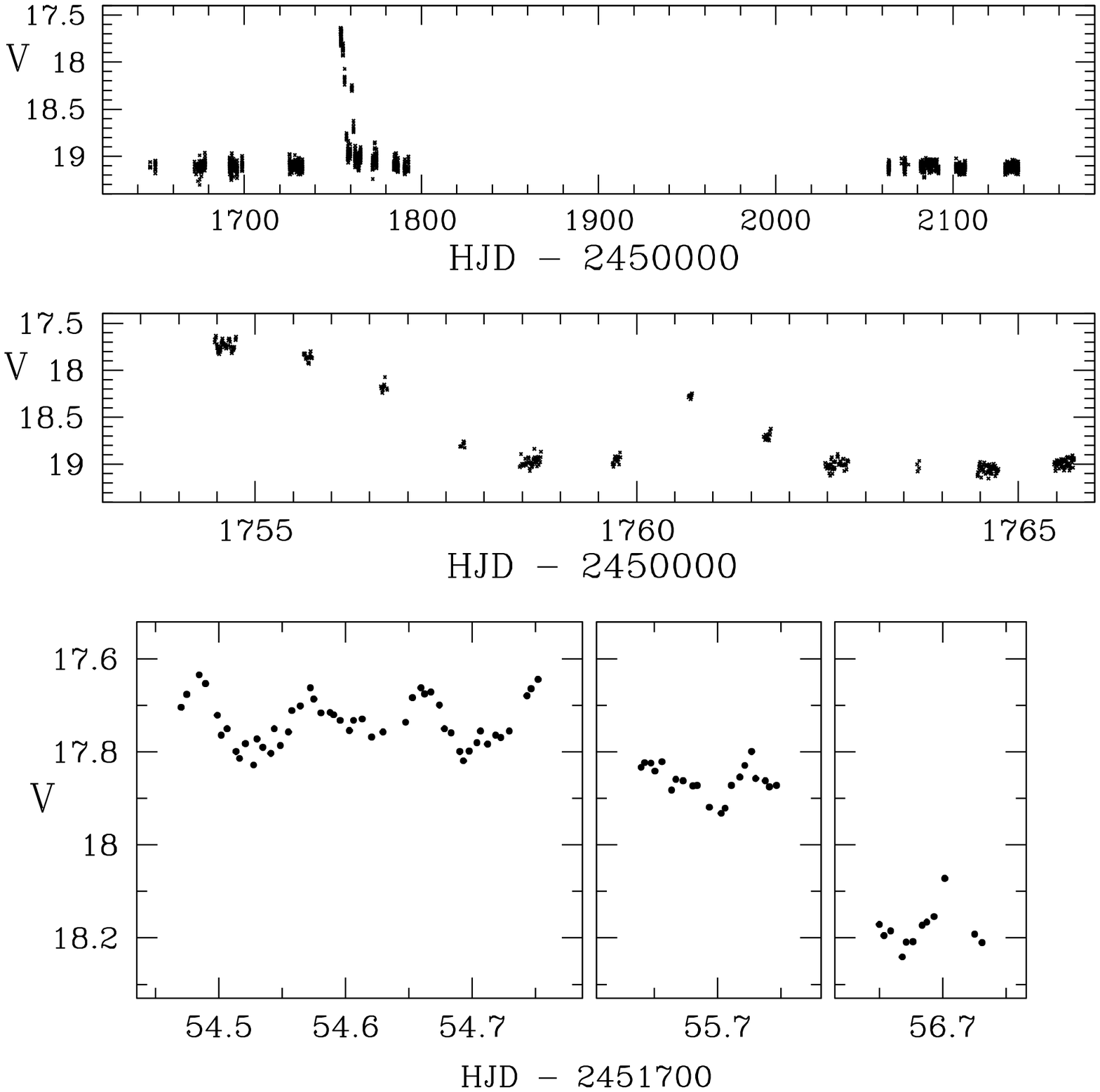}}
\FigCap{
Light curves of CV2: full span in the 2000-2001 seasons
(upper panel), during the 2000 July/August eruption (middle panel) 
and nightly segmented during the superoutburst (lower panels).
Note the presence of an ''echo'' outburst just after the superoutburst.
}
\end{figure}

\begin{figure}[htb]
\centerline{\includegraphics[height=120mm,width=120mm]{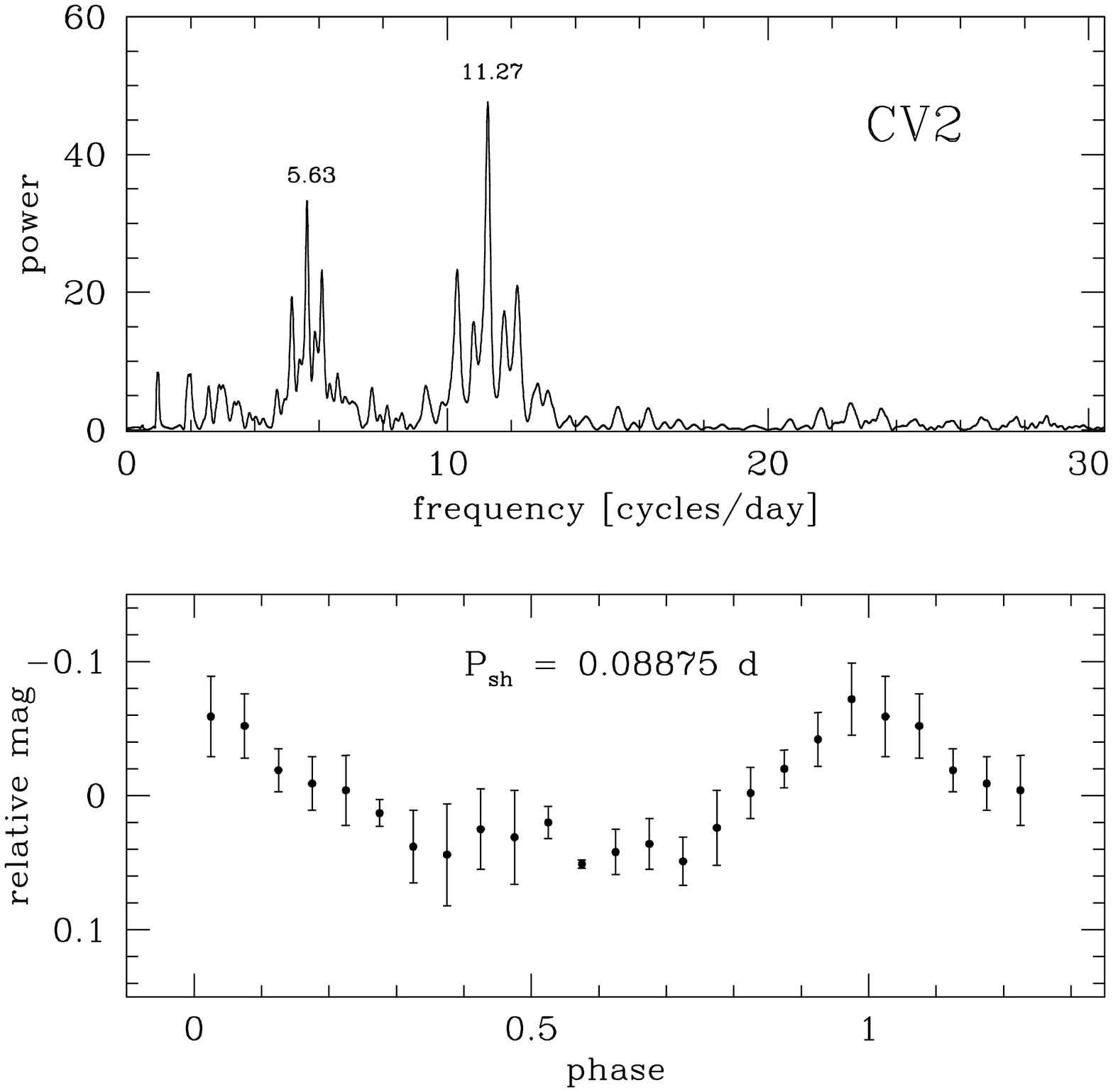}}
\FigCap{
Upper panel: the ANOVA power spectrum of CV2
of the JD=2451754-56 light curve of CV2. Lower panel:
the light curve folded at frequency 11.267~cycles/day.
}
\end{figure}

\begin{figure}[htb]
\centerline{\includegraphics[height=39mm,width=80mm]{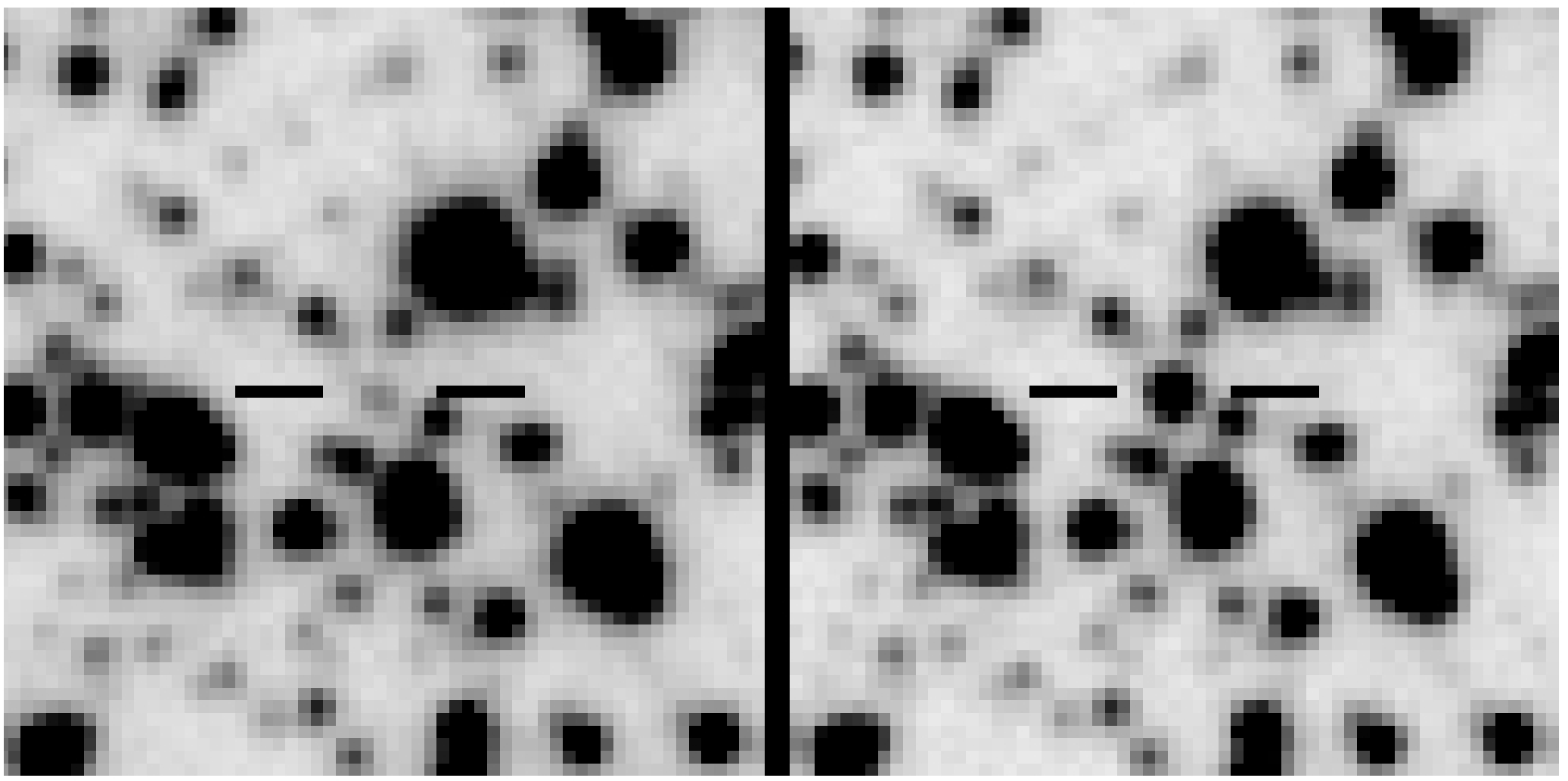}}
\FigCap{
Finding charts for CV2 showing the variable
in quiescence (left) and during outburst (right). North is up and
East is to the left. The field of view is $13\arcs$ on a side.
}
\end{figure}

Fig. 3 shows the light curve of the newly detected candidate DN.
A fading branch of an apparent outburst 
was observed during the period 2000 July 28-August 3. During that time
the brightness declined from $V\approx17.7$ mag to $V\approx 19.0$ mag.
The main outburst was followed by an "echo" eruption lasting 
about two days.
The light curve for the first three nights of observations covering 
the outburst (lower panel in Fig. 3) shows the presence of
superhumps. Their shape and full amplitudes of $\approx0.18$ mag 
are typical for SU UMa type DN (Warner 1985). 
After subtracting nightly means from 
the light curve we calculated the power spectrum with ANOVA statistics
(Schwarzenberg-Czerny 1996). The first two harmonics of the Fourier 
series were used in the fit.
The resulting periodogram is shown in the upper panel of Fig. 4.
The highest peak corresponds to a superhump period of
$P_{sh}=0.08875 \pm 0.00012$~days ($127.80 \pm 0.18$~min),
while the second peak is its subharmonic. The lower panel of
Fig. 4 presents the light curve folded with $P_{sh}$ and averaged  
in 0.05 phase bins. The orbital period of CV2 can be estimated using 
the empirical relation between $P_{sh}$ and
$P_{orb}$ first noted by Stolz and Schoembs (1984). Using the calibration
derived by Skillman and Patterson (1993) we obtain 
$P_{orb}=0.08496$~d (122.3~min). This places the variable at the 
lower edge of the gap observed in the distribution of orbital periods 
for CVs (Ritter and Kolb 2003).

Figure 5 presents finding charts for CV2 showing the variable 
in quiescence and in outburst. The variable is plotted on the cluster
color-magnitude diagram in Fig. 2. We derive $B-V=0.95 \pm 0.05$ mag
and $B-V=0.45 \pm 0.05$ mag for the low state and at maximum light,
respectively. Assuming a reddening of $E(B-V)=0.38\pm 0.04$
(Richter \etal 1999) we estimate the unreddened color
at maximum light to be $(B-V)_{0}\approx 0.07$ mag.
Most DNe at maximum light have unreddened colors in the range 
$(B-V)_{0}=0.0\pm 0.10$ mag (Warner 1976). 
The variable was observed at $V\approx 19.0$ mag in quiescence.
This brightness is consistent with the hypothesis that CV2 belongs to M22,
despite being located at 3.9 core radii from the cluster center.
Assuming cluster membership
the variable has $M_{V}\approx 5.5$ mag in quiescence (we
adopt $(m-M)_{V}=13.5$ mag for M22, Harris 1996). This places CV2 in the 
range of absolute magnitudes observed for bright non-magnetic DN.

Further confirmation of the cataclysmic nature of CV2 comes from the 
X-ray observations of M22 obtained with XMM-Newton.
Webb \etal (2004) published a list of 50 X-ray sources detected
in the $30' \times 30'$ field centered on the cluster. One of the sources,
object \#40, is a very probable counterpart of CV2. 
The  equatorial coordinates of the X-ray object are 
$\alpha_{2000}=18^h36^m02\zdot\ups96$,
$\delta_{2000}=-23^{\arcd}55^{\arcm}26\zdot\arcs42$,
offset from the optical coordinates
of CV2 listed in Table 1 by ($\Delta \alpha=3\zdot\arcs6$,
$\Delta \delta=1\zdot\arcs8$), well within the $5\arcs$
uncertainty circle of the X-ray source position. The XMM-Newton observations
were obtained on Sep 19-20, 2000. That was 20~days after our last observation
in the 2000 season and 45~days after the eruption of CV2.
The measured count rate in the 0.5-10.0~keV band for source \#40 was 2.74
times lower than that measured for source \#36 associated with CV1. 
For CV1 Webb \etal (2004) give an unabsorbed flux of
$9.8 \times 10^{-14}$~erg~cm$^{-2}$s$^{-1}$.
Assuming that both CVs belong to the cluster at a distance of
3.2~kpc (Monaco \etal 2004) we can estimate the X-ray 
luminosity of CV2 during quiescence as
$L_X\approx4.4 \times 10^{31}$~erg~s$^{-1}$. This value is consistent
with X-ray luminosities observed for field non-magnetic CVs
(\eg Baskill \etal 2005).

\Section{Microlensing Event in M22}

Our search for erupting objects in M22 also led to the detection of a
source which underwent one episode of increased luminosity. Its light
curve is shown in Fig. 6. The source is located 140$^{\arcs}$ from
the cluster center in a relatively crowded region (see the chart in
Fig. 7). The brightness of the object increased by $\approx0.7$ mag
over 20 days. Around 2000 August 5 it reached a maximum of
$V\approx19.1$ and then faded to a constant level of $V\approx19.9$ mag.
On the cluster color-magnitude diagram (Fig. 2) the star is placed
among the cluster main sequence stars. However, we note that at minimum
brightness it is only marginally detectable on $B$ images leading to a
large uncertainty in the measured color. 
The variable is relatively red at the at the maximum light
with $B-V=0.97\pm 0.09$ mag and $(B-V)_0\approx 0.59$ mag. \footnote{
According to Schlegel \etal (1998) the total reddening in the 
cluster direction amounts to $E(B-V)=0.33$ mag. That implies
that any stars located in the cluster field 
cannot be significantly more reddened than the cluster itself.}
This excluded possibility
that the object belongs to DNs (Warner 1976). At minimum light we
measured $B-V=1.24\pm 0.22$ mag what is consistent with a lack of any 
color change during an outburst. 

\begin{figure}[htb]
\centerline{\includegraphics[height=120mm,width=120mm]{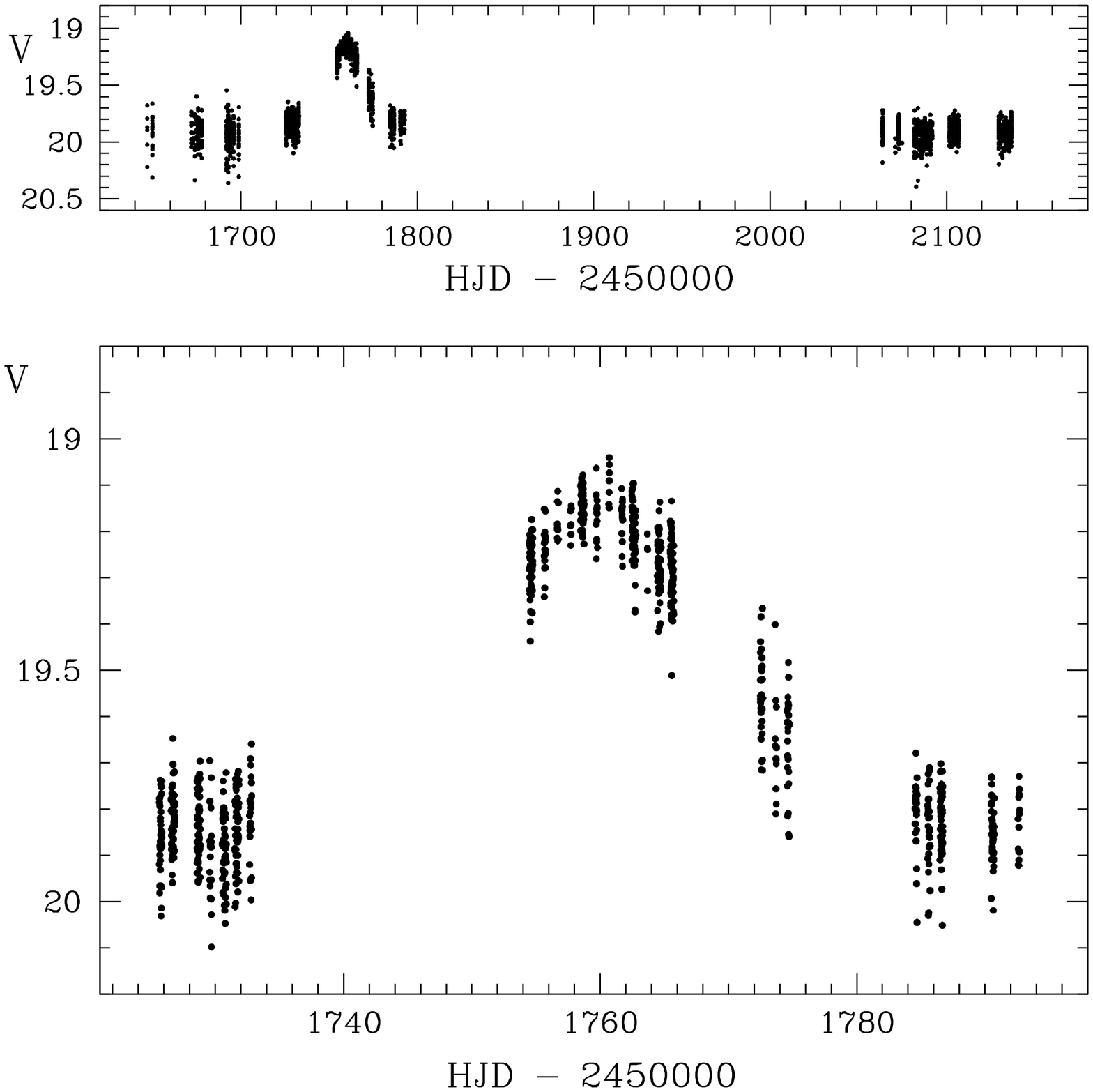}}
\FigCap{
Full span (top) and event (bottom) light curves
of the probable microlens in M22.
}
\end{figure}

\begin{figure}[htb]
\centerline{\includegraphics[height=39mm,width=39mm]{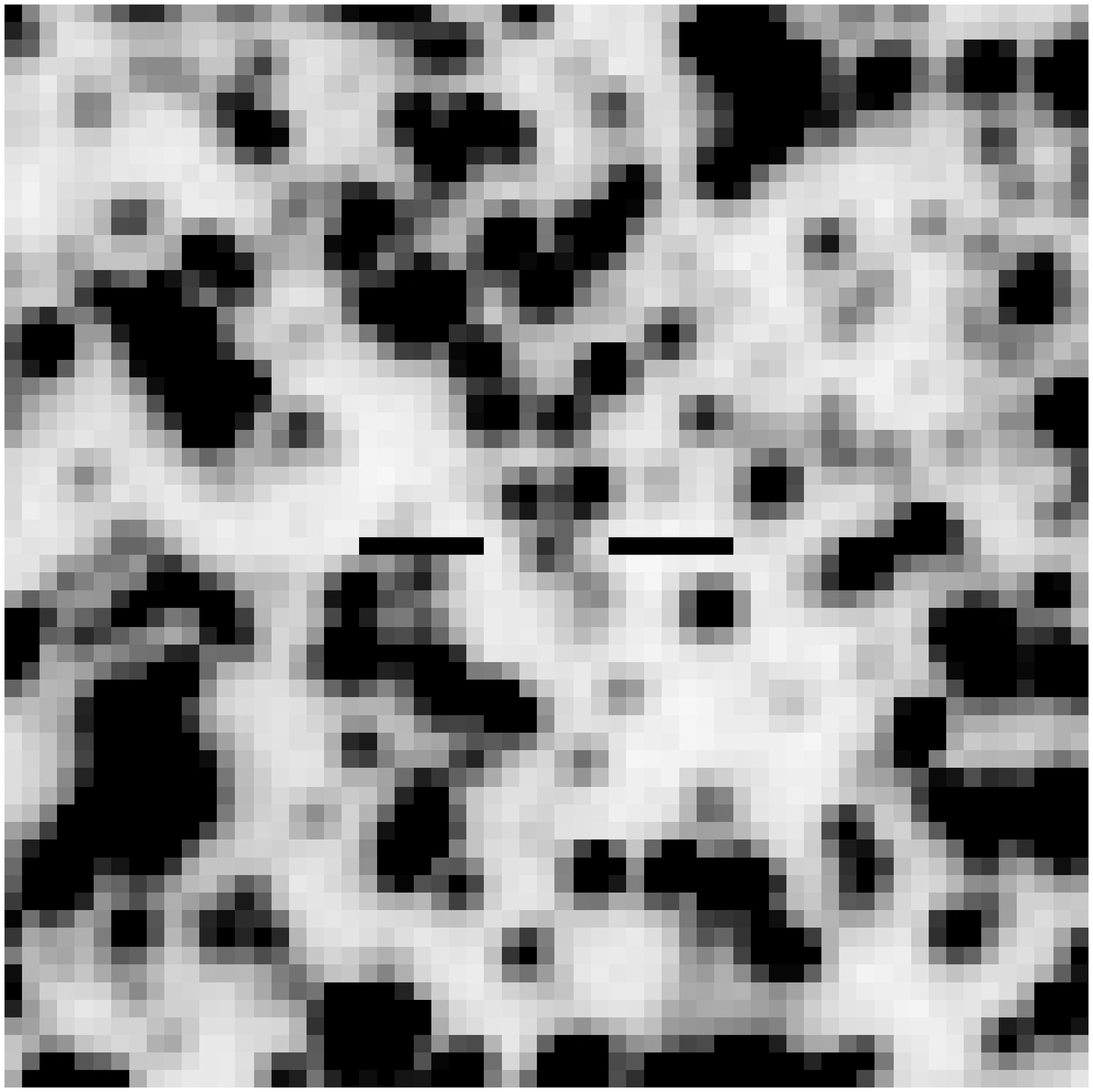}}
\FigCap{
Finding chart for the probable microlens in M22.
North is up and East is to the left. The field of view is $13~\arcs$
on a side.
}
\end{figure}

The observed behaviour of the star is characteristic  
of a gravitational lens event. To check this hypothesis we use a
5-parameter model of a single lens event and fit a light curve of the
standard form (Paczy\'nski 1986):

$$
F(t)=A(u(t))~F_{\rm s} + F_{\rm b}
$$
\begin{equation}
A(u)=\frac{u^2+2}{u\sqrt{u^2+4}} 
~~~~u(t)=\sqrt{u_0^2+\frac{(t-t_0)^2}{t^2_{\rm E}}}
\end{equation}

where $F_{\rm s}$ is the (unmagnified) flux of the star undergoing
lensing, $F_{\rm b}$ - the blended flux, $t_0$ - the epoch of
maximum, $t_{\rm E}$ - a characteristic ({\it Einstein}) time of the
event, and $u_0$ the impact parameter expressed in units of the Einstein
radius. The Einstein radius for a lens of mass $M$ located at a distance
$d_{\rm L}$, and a source at a larger distance $d_{\rm S}$ is:

\begin{equation}
r_{\rm E}=\sqrt{\frac{4GM}{c^2}
\frac{d_{\rm L}(d_{\rm S}-d_{\rm L})}{d_{\rm S}}}
\end{equation}

where $G$ is the gravity constant and $c$ the speed of light.

We transform the measured stellar magnitudes $V_i$ 
and their error estimates $\Delta V_i$ into
flux units  ($F_i$, $\sigma_i$) and fit the model minimizing $\chi^2$:

\begin{equation}
\chi^2=\sum_{i=1}^N\frac{\left(A_i~F_s+F_b-F_i\right)^2}{\sigma_i^2}
\end{equation}

where $A_i \equiv A(u(t_i))$. 

The fitted parameters, with source and blend fluxes 
converted back into stellar magnitudes $V_{\rm s}$, $V_{\rm b}$ are:

$$
t_0=2451759.70^{+0.33}_{-0.34}~~~~
t_{\rm E} = 15.^{\rm d}9^{+5.0}_{-1.1}~~~~
u_0=0.54^{+0.02}_{-0.18}
$$
$$
V_{\rm s}=19.92^{+0.62}_{-0.02}
~~~~~~V_{\rm b}=24.8^{+\infty}_{-4.0}
$$
$$
\chi^2=127~~~~~~{\rm DOF}=1923
$$

The fit, shown in Fig. 8 is not well constrained due to 
the Wo\'zniak and Paczy\'nski
(1997) degeneracy. The fitted parameters are highly correlated within the
confidence regions in parameter space. The event can be fitted by
a relatively shorter lasting and weaker amplification of a brighter source with
a fainter blend as well as by a longer lasting and stronger amplification
of a fainter source with a brighter blend.  

\begin{figure}[htb]
\centerline{\includegraphics[height=120mm,width=120mm]{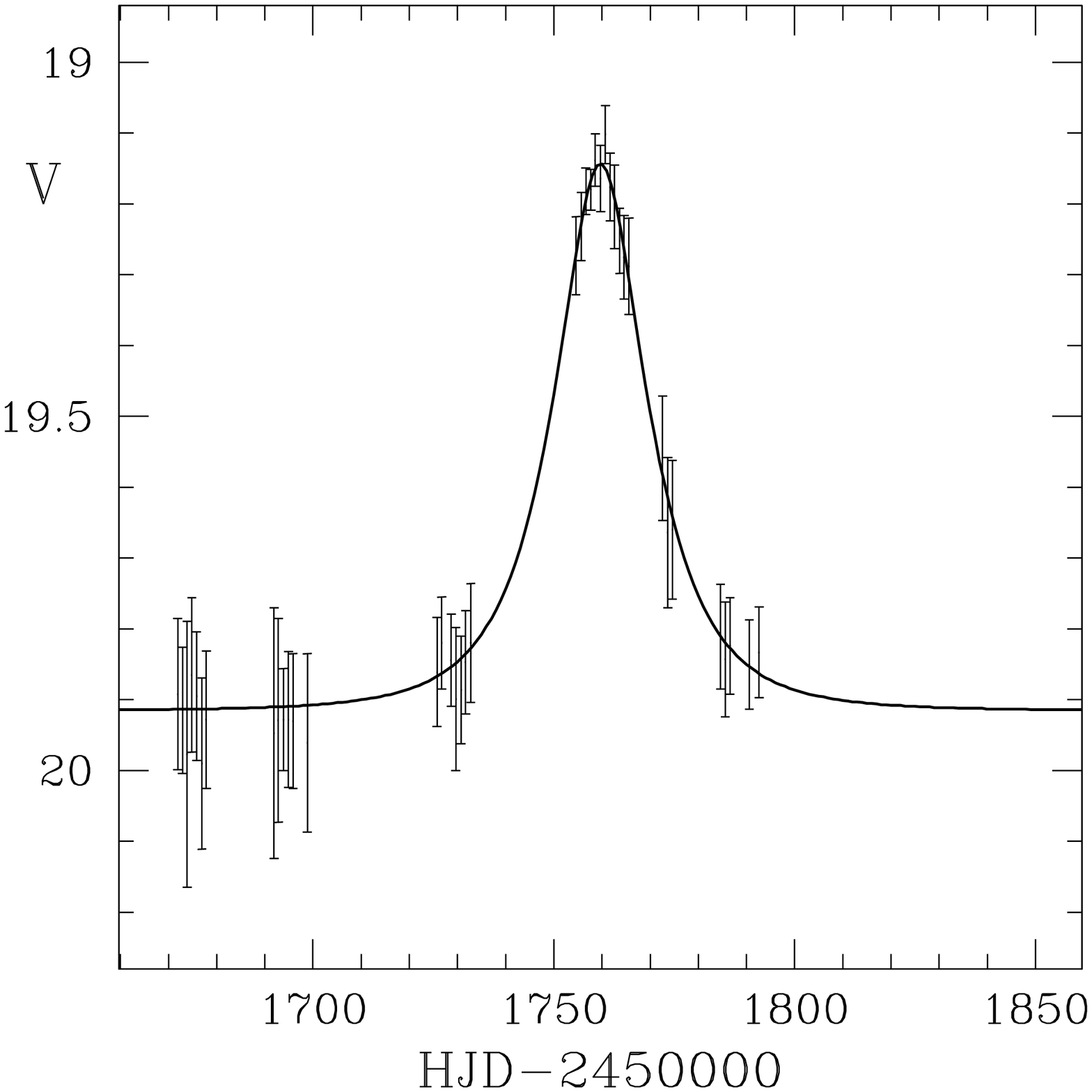}}
\FigCap{
The standard (5 parameter) binary lens model of the
microlensing event. The data points were nightly binned for clarity
of the plot.
}
\end{figure}

The source and the lens can be located in the bulge, in M22, or anywhere
along the line of sight. Since the large majority of objects of interest
belong either to M22 or to the bulge, we shall consider primarily these
locations. First we assume that the lens belongs to the cluster and
compare the {\it a priori} relative probability of the source being
in the bulge or in the cluster. 

We adopt the distances $d_{\rm M22}=3.2{\rm kpc}$ to the cluster
(Monaco \etal 2004)
and $d_{\rm B}=8.0{\rm kpc}$ to the bulge. The Einstein ring
projected into the source plane has a radius
${\tilde r}_{\rm E}=r_{\rm E}~d_{\rm S}/d_{\rm L}$. Only 
sources within ${\tilde r}_{\rm E}$ from the observer - lens
line are viable candidates for lensing.

For a rough estimation of the relative lensing probability we adopt a
simplified model of the cluster assuming spherical symmetry and a
density profile $n(r) \sim r^{-3}$ outside the core. The line of sight
to the lensed source passes at $b_{\rm M22} = 2.17{\rm pc}$ from the 
cluster center while the core radius $r_{\rm c}=1.30{\rm pc}$
(Trager \etal 1995). In our model the surface density of cluster stars
is related to its space density along the line of sight:

\begin{equation}
N_{\rm M22}=
\int_{-\infty}^{+\infty}~n_0\frac{b^3}{(b^2+x^2)^{3/2}}~dx=2n_0 b
\end{equation}

where $n_0$ is the 3D star density at a distance $b$ from the cluster
center. The probability of finding a cluster source within 
the Einstein radius from a lens of mass $M$ is:

\begin{eqnarray}
P_{\rm M22}=
\frac{1}{N_{\rm M22}}
\int_{-\infty}^{+\infty}dx~n(x)~\int_x^{+\infty}dy \bigg\{n(y)~
\pi\frac{4GM}{c^2} \times {}
\nonumber\\
{} \times \frac{(d_{\rm M22}+y)(y-x)}{d_{\rm M22}+x} \bigg\} 
\approx N_{\rm M22}\frac{2\pi GM}{c^2}b_{\rm M22} 
\end{eqnarray}

where $x$ and $y$ are the lens and source positions respectively,
measured along the line of sight.  
We approximate $(d_{\rm M22}+y)/(d_{\rm M22}+x) \approx 1$ and
then solve the double integral analytically.

For a source in the bulge the lens position within the cluster is
irrelevant, and one gets a probability of lensing:

\begin{equation}
P_{\rm B}= N_{\rm B}\frac{4\pi GM}{c^2}
\frac{d_{\rm B}(d_{\rm B}-d_{\rm M22})}{d_{\rm M22}}
\end{equation}

The relative probability of the lensed source belonging to the cluster is
$P_{\rm M22}/P_{\rm B}=9.04 \times 10^{-5}N_{\rm M22}/N_{\rm B}
\approx 5 \times 10^{-4}$ where we use estimates of the surface
densities of stars belonging to M22 and the bulge from Albrow \etal (2002).

Similarly, assuming that the source is located in the bulge, one can
check the relative probability of finding the lens in the bulge or in
the cluster. The calculations are analogous to those presented above, but the
surface density of sources should be replaced by surface mass density of
lenses. The line of sight passes at $\approx 12^\circ$ or
$b_{\rm B} \approx 1.7{\rm kpc}$ from the bulge center.
Calculations give

\begin{equation}
P_{\rm B}=\frac{2\pi G\Sigma_{\rm B}}{c^2}b_{\rm B}
\end{equation}
\begin{equation}
P_{\rm M22}=\frac{4\pi G\Sigma_{\rm M22}}{c^2}
\frac{d_{\rm M22}(d_{\rm B}-d_{\rm M22})}{d_{\rm B}}
\end{equation}

where $\Sigma_{\rm B}$, $\Sigma_{\rm M22}$ stand for the surface
mass density in the bulge and in M22, respectively, measured along the
line of sight.
The geometrical factors are of the same order, so 
$P_{\rm B}/P_{\rm M22}=0.42\Sigma_{\rm B}/\Sigma_{\rm M22}$.
Assuming that surface mass densities are proportional to the surface
densities of stars we get $P_{\rm B}/P_{\rm M22} \approx 0.09$.
The models of mass distribution in M22 and bulge suggest even an smaller
value. Thus our estimates show that the location of the source in the
bulge with lens in M22 is the most likely situation. The location
of both objects in the bulge is $\ge 11$ times less likely, while the
location of both objects in M22 is $\approx 2000$ times less likely.
In the following we assume that the source is in the bulge and the lens
in the cluster.

The proper motion of the lens relative to the source is dominated by the
motion of the cluster as a whole (Gaudi 2002), 
$\mu_{\rm rel}=10.9~{\rm mas}~{\rm y}^{-1}$ (Peterson and
Cudworth 1994). The velocity dispersion of stars in M22, 
$\sigma =11.4~{\rm km}~{\rm s}^{-1}$ (ibid.) corresponds to 
$\mu = 0.75~{\rm mas}~{\rm y}^{-1}$, and the motion of stars in the
bulge with velocity dispersion components 
$(\sigma_{\rm l},\sigma_{\rm b})=(93,79)~{\rm km}~{\rm s}^{-1}$
(Han and Gould 1996) corresponds to 
$\mu \approx 2~{\rm mas}~{\rm y}^{-1}$, so they can be neglected
to simplify the analysis. 
By the definition of Einstein time we have 
$r_{\rm E}=d_{\rm M22}\mu_{\rm rel}t_{\rm E}$, which leads
to the following mass estimate for the lens:

$$
M=0.14~M_\odot~\left(\frac{t_{\rm E}}{15.9~{\rm d}}\right)^2
=0.14^{+0.10}_{-0.02}~M_\odot
\eqno(10)
$$

where we neglect all sources of error except the uncertainty in the
Einstein time fit.

\Section{Discussion and Summary}

We have presented the results of a search for erupting
objects in the field of the globular cluster M22. 
A new cataclysmic variable, CV2, 
which underwent a DN type superoutburst in 2000 was detected.
Prominent superhumps with a period of 128 minutes
were observed in the light curve during three nights.
CV2 has an X-ray counterpart detected by the XMM-Newton telescope.
The cluster membership of the object remains an open issue.
Recent dynamical models of GCs (Heggie and Hut 2003)
predict that close binaries should sink to the centers of the clusters
due to encounters with neighboring stars.
CV2 is located at a distance of 3.9 core radii from the center
of M22. However, we note that there are several examples of 
close binaries located in outskirts of their parent  
clusters (eg. Bassa \etal 2003; Margon \etal 1981). 
The position of CV2 on the color-magnitude 
diagram of M22 and its observed X-ray luminosity 
are both consistent with the hypothesis that the variable is a cluster member.
This issue can be easily settled by measuring the systemic radial 
velocity of the variable. The cluster itself has $V_{rad}=-149$~km/s
which differentiates its stars from the background population. 
We also detected two DN outbursts for the previously identified variable CV1. 

The search for erupting objects also led to the detection of
a probable microlensing event in M22. It is located 2.3 arcmin
from the cluster center and had an amplitude of 0.75 mag. We fitted
the light curve to a 5-parameter model of a single lens event.
The most likely geometry of the event places the source in the 
Galactic bulge and the lens in the cluster.

\Acknow{

PP would like to thank Arek Olech for helpful comments
during data analysis. This work was supported by the following grants
from the Ministry of Science and Information Society Technologies, Poland:
1~P03D~024~26 (to PP), 1~P03D~001~28 (to JK) and 2~P03D 016 24 (to MJ).
}

\end{document}